\documentclass[amsmath,twocolumn,amssymb,prd,preprintnumbers,showpacs]{revtex4-1}
\usepackage{amsmath}
\usepackage[cmtip,arrow]{xy}
\usepackage{pb-diagram,pb-xy}
\usepackage{slashed}
\usepackage{amssymb,amsopn}
\def \be {\begin{equation}}
\def \ee {\end{equation}}
\def \bea {\begin{eqnarray}}
\def \eea {\end{eqnarray}}
\def \sla {\slashed}
\usepackage{graphicx}
\begin{document}
\title{Myers-Pospelov Model as an Ensemble of 
Pais-Uhlenbeck Oscillators: 
Unitarity and Lorentz Invariance Violation}
\author{Justo Lopez-Sarrion}
\affiliation{Departamento de F{\'i}sica, Universidad 
de Santiago de Chile, Casilla 307, 
Santiago, Chile.}
\author{Carlos M. Reyes}
\affiliation{Departamento de Ciencias B{\'a}sicas, 
Universidad del B{\'i}o B{\'i}o, Casilla 447, 
Chill\'an, Chile.}
\begin{abstract}


{We study a generalization of a Pais-Uhlenbeck oscillator 
for fermionic
variables. Next, we consider an ensemble of these 
oscillators and we
identify a particular case of the Myers-Pospelov 
model which is
relevant for effective theories of quantum gravity. 
Finally, by taking the  advantage of this connection, we analyze, for this model, the unitarity at one loop order in the low energy regime where no ghost states can be created on-shell. This energy regime is the relevant one when we consider the Myers-Pospelov model as a true  effective theory coming from new space-time structure.}
\end{abstract}
\maketitle
\section{Introduction}
In recent years, there has been growing interest 
in quantum field 
theories containing higher time derivatives. 
In part because they arise naturally as effective corrections  
from candidate fundamental theories such as 
string theory \cite{Strings}, 
quantum gravity \cite{Stelle} 
and noncommutative geometry \cite{NC-Geom}.
Moreover, attempts to 
incorporate such terms 
have also been given in the context of dark energy \cite{D-E}, 
Lorentz invariance
violation \cite{L-V,MP,BP}, radiative corrections \cite{Radiative,Mariz}
perturbative
iteration \cite{Jaen} and 
regularization \cite{Regularization}.

Many years ago, a quantum mechanical system with higher time 
derivatives was proposed by Pais and Uhlenbeck (P-U)
\cite{P-U}. This system provides a generalization of the 
harmonic oscillator by including a fourth time derivative.
The first problem that one faces is the unbounded 
character of the spectrum \cite{Ostro}; sometimes, 
however, constrained systems can 
turn around the problem, see \cite{M-P} and 
there has been also other proposals \cite{Other}. 
There are some redefinitions 
which
avoid these problems but in turn they introduce states 
of negative norm. Lee and Wick (L-W) generalize these 
ideas to relativistic
quantum field theories \cite{INDFMETRIC}. These 
undefined metric theories 
have the advantage of being finite in the sense that these 
higher time derivatives regularize the 
divergences. The problem of unitarity, which comes out 
with undefined
metric, could be controlled by a prescription due to 
Cutkosky \cite{CUTW}.
Recently, these models have been used as an extension 
of the standard 
model of particles which fixes the hierarchy problem 
\cite{BG}. However, models incorporating operators 
of mass
dimension 5, which violate Lorentz symmetry, 
might also contain higher time derivatives. 
These theories due to Myers and Pospelov (M-P) \cite{MP} are 
supposed to arise from quantum gravity effects at 
high energies. These two last examples,
are part of the increasing interest on higher
 time derivatives in quantum field theory and, 
 therefore, it is compelling 
to focus on the technical and physical issues 
concerning these kind of theories.

In this work, we propose to construct a field 
theory as ensembles of Pais-Uhlenbeck oscillators
in momentum space, instead of harmonic 
oscillators of the standard relativistic case. 
We find that this new theory is close 
related to the Myers-Pospelov fermionic sector.
Taking advantage of this relation, 
we explore the unitarity of the model by modifying 
the Cutkosky prescription.

All the theories with higher time derivatives seem to have an unavoidable problem because they content negative norm states, at least if there is no gauge symmetry  that prevents these modes from appearing in the spectrum \cite{M-P}. The fermonic sector of the Myers-Pospelov model also has this feature, and these ghost modes appear even at tree level in perturbative  theory, when the initial state have a total energy greater than the mass of one of these modes. However, this model  has to be seen as an effective  model, only  valid for energies  below certain scale, beyond which new physics  is described through a more fundamental theory. Hence, lack of unitarity and probabilistic interpretation problems coming  from  the ghost states are irrelevant because the description of physics beyond  that threshold with this model is not justified.  Nevertheless, the model must be well defined  below  the threshold, and it must be capable of giving sensible predictions at this  low energy regime. But, unitarity could be spoiled out due to virtual processes even though no on-shell ghosts can be created at low enough energies. This problem was addressed by Cutkosky {\it et al.}\cite{CUTW} in the context of relativistic Lee-Wick theories, as mentioned above. They used a convenient prescription to define the propagators in order to keep unitarity order by order  in the loop expansion. We will follow the same  philosophy to  define the propagators in such a way that the Myers-Pospelov model preserves unitarity below the threshold scale. It is worth noting that in the work by Cutkosky the Lorentz invariance is and important ingredient to be considered. However, in our  present case, relativistic  symmetry only plays a secondary role which will serve to fix part of the prescriptions which connects to the standard relativistic theory at very low energies compared with the threshold.

The layout of this work is the following. 
In the second section, we review the 
Pais-Uhlenbeck oscillator
and we generalize it to fermions. In 
the third section, we construct an ensemble 
of P-U fermionic 
oscillator and we identify it as the fermion 
sector of M-P model. In section four, we analyze the unitarity of the Myers-Pospelov model at one loop order,  in the low energy regime by using the structure of P-U oscillators.
Finally, we give the conclusions and final 
comments.\section{The issue of higher time derivatives}
In this section, we introduce the P-U model and we 
generalize it to the fermionic case
which may be less familiar, and hence it will be studied with more details.
\subsection{The Pais-Uhlenbeck model}
The P-U oscillator is, basically, the standard harmonic 
oscillator with an additional higher time
derivative term. To be more precise its equation of motion is 
\begin{eqnarray}
 g q^{(4)}+\ddot q+\omega^2q=0,
\end{eqnarray}
where $q^{(4)}$ is a fourth order time derivative and $g$ 
can be considered a small coupling constant.
The equation of motion is obtained from the Lagrangian: 
\begin{eqnarray}
L_{PU}= -\frac{g}{2} \ddot q^2+\frac{1}{2} \dot q^2-\frac{1}{2}\omega^2q^2.
\end{eqnarray}
This system can be seen as two standard harmonic oscillators 
by means of the change of variables, 
\begin{eqnarray}
q_+&=&(\partial_t^2-k_{-}^2) q,\nonumber  \\ q_-&=&(\partial_t^ 2-k_{+}^2) q.
\end{eqnarray}
The Lagrangian with the new variables is 
\begin{eqnarray}
L_{PU}= \frac{1}{2} \dot q_+^2-\frac{1}{2}k_{+}^2q_+^2-
\frac{1}{2} \dot q_{-}^2+\frac{1}{2}k_{-}^2q_{-}^2,
\end{eqnarray}
with $k^2_{\pm}= \frac{1}{2g} (1\mp \sqrt{1-4g\omega^2})$ 
positive frequencies depending on $g$ and $\omega$.
 This Lagrangian corresponds
to two standard harmonic oscillators with one of them 
having a relative minus sign 
respect to the other. This last fact makes a great
 difference with the simple sum of two oscillators
when we add interactions or quantize the theory. 
Classically, the equations of motion are just
the same as two standard harmonic oscillators, 
however when adding interactions the system 
becomes unstable due to 
the unboundedness of the energy. This problem 
persists under the usual quantization. However,
by a redefinition of the vacuum state, the quantum 
mechanical problem become stable.
This last procedure leads to unavoidable negative norm 
states. These ghosts states could render
the theory non unitary when interaction are considered. 
To be more precise, following the canonical formalism 
we write the Hamiltonian as:
\begin{eqnarray}
\hat H_{PU}=k_{+}\hat a_{+}^{\dag}\hat a_{+}- k_{-} \hat 
a_{-}^{\dag}\hat a_{-}+\frac{1}{2}(k_{+}-k_{-}),
\end{eqnarray}
where
$\hat a_{+}$, $\hat a_{+}^{\dag}$, $\hat a_{-}$, $\hat 
a_{-}^{\dag}$ are the standard creation and
annihilation operators. 
The second term 
produces arbitrary negative energy states as 
can be seen 
by acting $\hat a_{-}^{\dag}$ on the empty 
wave function (defined by
$\hat a_{+} \Phi_0=\hat a_{-}\Phi_0=0$)
\begin{eqnarray}
\Phi_0 &=&N \exp \left[-\frac{\sqrt{1-4g\omega^2}}
{2(k_{+}+k_{-})} (k_{+} k_{-}q^2+\dot q^2 ) + \sqrt{-g\omega^2}q
\dot q\right]. \nonumber
\\
& &
\end{eqnarray}
An alternative proposal to quantize would be to
 redefine the vacuum $\hat a_+ \Phi_0'=\hat a_{-}^
 {\dag} \Phi_0'=0$.
Then the states $\psi^{'}_{n}=a_{-}^{(n)} 
\psi^{'}_0$
 have positive energy, but are ghost states 
 when $n$ is odd. 
Hence, this procedure stabilizes the theory 
but it 
could spoil unitarity when interactions are 
introduced.
\subsection{Fermionic extension of the P-U model}
Now we consider the fermionic version of the 
previous model with equation of motion:
\begin{equation}\label{eqmotion}
g\ddot\psi+i\dot\psi-\omega\psi = 0.
\end{equation}
As before, this is the fermionic harmonic 
oscillator ($g=0$) plus
a higher derivative term. Without loss of generality we can assume 
that $\omega$ and $g$ are positive constants.

This equation of motion comes from the Lagrangian 
\begin{eqnarray}\label{lagori}
L_{F}= g\bar\psi\ddot\psi + i\bar\psi\dot\psi -\omega\bar\psi\psi.
\end{eqnarray}
Let us consider the new variables
\begin{eqnarray}\label{def1}
\psi_{-}&=& \alpha(i\partial_t - \omega_+ )\psi,\nonumber 
\\ \psi_{-}^\dag&=& \alpha(-i\partial_t -\omega_+){\bar\psi}\label{c-},
\end{eqnarray}
and
\begin{eqnarray}\label{def2}
\psi_{+}&=& \alpha(i\partial_t - \omega_-)\psi,\nonumber \\
\psi_+^{\dag} &=& \alpha(-i\partial_t - \omega_-)\bar \psi\label{c+},
\end{eqnarray}
where $\alpha=\left(\frac{g}{\omega_{-}-\omega_{+}}
\right)^{\frac{1}{2}}$ and the positive frequencies $\omega_{\pm}$ are defined by
\begin{eqnarray}
\omega_\pm = \frac{1\mp \sqrt{1-4g\omega}}{2g},
\label{omega}
\end{eqnarray}
where $\omega_{-}>\omega_{+}>0$.
The Lagrangian in terms of this new variables is
\begin{eqnarray}
L=  \psi_{+}^{\dag}  (i\partial_t-\omega_+)   \psi_{+} -  
\psi_{-}^{\dag}  (i\partial_t-\omega_{-})   \psi_{-},  
\end{eqnarray}
which can be reduced to the original Lagrangian (\ref{lagori}), by
replacing Eqs. (\ref{def1}) and (\ref{def2}) on it.

Now, the usual variation of the above Lagrangian 
gives the equations of motion 
\begin{eqnarray}\label{eqmotiondec}
  (i\partial_t-\omega_+)   \psi_{+}&=&0\nonumber ,\\ 
  (i\partial_t-\omega_{-})   \psi_{-} &=&0.
\end{eqnarray}
As in the bosonic case the system is decoupled into two 
standard oscillators, but with a global minus sign for the $\psi_{-}$. 
Following the canonical quantization for fermions, one can check that the canonical
conjugate momentum for $\psi_+$ is $i \psi_+^{\dag}$, but for $\psi_{-}$
is $-i\psi_{-}^{\dag}$. Hence, this gives the anticommutators in the 
new variables 
\begin{eqnarray}\label{anticomm}
\{\psi_{+},\psi_{+}^\dag\}&=& 1,\nonumber \\
 \{\psi_{-},\psi_{-}^\dag\}&=& -1.
\end{eqnarray}
Then, the operators $\psi_-$, $\psi_-^\dag$, $\psi_+$ 
and $\psi_+^\dag$ are standard 
creation and annihilation operators, respectively, but $\psi_{-}$ and $\psi_{-}^\dag$ create 
and annihilate states of negative norm.

The original variables can be written in terms of 
these operators as
\begin{eqnarray}
\psi&=&\frac{\psi_{-}-\psi_+}{(1-4g\omega)^{1/4}},
\label{pisc}\nonumber \\
\dot\psi &=&-i\frac{\omega_- \psi_- - \omega_+ \psi_+}
{({1-4g\omega})^{1/4}}, \label{psicdot}
\end{eqnarray}
so the Hamiltonian turns out to be 
\begin{eqnarray}
H_F = \omega_{+} \psi_{+}^\dag \psi_{+}  - \omega_{-} 
\psi_{-}^\dag \psi_{-} .
\end{eqnarray}
This expression can be compared to the one 
for the P-U 
Hamiltonian. They are the same expression, 
except for an irrelevant constant. 
However, this system corresponds to a four state 
system, and hence, the spectrum is bounded 
below, and the vacuum is 
normalizable. In fact, the true vacuum is 
the state annihilated by 
$\psi_-$ and $\psi_+$, namely, $\psi_-\Phi_0=\psi_+\Phi_0=0$
where the explicit form of the wave function for the 
vacuum has been calculated 
in the appendix. 

Moreover, the rest of the 
states are $\Phi_1=\psi_+^\dag\Phi_0$, with energy $\omega_+$,
$ \Phi_2=\psi^\dag_-\Phi_0$ with energy $\omega_{-}$
and 
$\Phi_3=\psi^\dag_-\psi_+^\dag \Phi_0$ with energy $\omega_{+}+\omega_{-}$. 
Nevertheless, the $\Phi_2$ and $\Phi_3$ operator correspond to ghost states.
In the appendix we follow the Schr\"{o}edinger 
quantization 
procedure and the interpretation is the same as above.

Therefore, in the fermionic case we just have one 
alternative to quantize the theory,
in contrast with the bosonic case where we can 
either have unstable spectrum or indefinite norms
in agreement with the analysis with Lee and Wick 
\cite{INDFMETRIC}.

Finally, we calculate the temporal ordered Green's 
functions which will be useful in later perturbative
considerations.
To do this, let us consider the usual definition
\begin{eqnarray}
G_{ab }^F(t)&=&\theta(t)\langle 0   \vert   
\psi_{a}(t)\psi^{\dag}_{b}(0) \vert 0\rangle -
\theta(-t)\langle 0   \vert \psi^{\dag}_{b}(0)
\psi_{a}(t)     \vert 0\rangle 
\nonumber \\ &=&  \langle 0   \vert    \mathcal{T}(\psi_{a}(t)
\psi^{\dag}_{b}(0))\vert 0\rangle, 
\end{eqnarray}
where $a,b=\pm$.
By solving the equations of motion (\ref{eqmotiondec}) we have
\begin{eqnarray}
\psi_{a}(t)=e^{-i\omega_{a}t} \psi_{a}(0).
\end{eqnarray}
Replacing this in the Green function expression and 
taking into account that 
$\psi_{a}(0)\vert 0 \rangle =0$ we arrive at
\begin{eqnarray}
G_{ab}^F(t)=a e^{-i\omega_{a} t}  \theta(t)\delta_{ab}, 
\end{eqnarray}
where we have made use of the anticommutation 
relations (\ref{anticomm}). This Green function 
satisfies the equation
\begin{eqnarray}
  (i\partial_t-\omega_{a})  G_{ab }^F(t)= a i \delta(t)\delta_{ab}. 
\end{eqnarray}
We can use an integral representation with the 
following prescription
\begin{eqnarray}\label{green}
G_{ab }^ F(t)=ai \delta_{ab}\int \frac{d \omega}{2\pi} 
\frac{e^{-i\omega t }}{\omega -\omega _{a }+i\epsilon}, 
\end{eqnarray}
where the prescription $\epsilon>0$ is used.

Note that we are dealing with a $(0+1)-$ 
dimension model where there is no 
particles and antiparticles. In a more 
realistic $(3+1)-$ dimension case, particles and 
antiparticles are present, and the usual prescription for 
$\epsilon$ is positive for particles and negative 
for antiparticles.
\section{Fermionic Pais-Uhlenbeck ensemble and the Myers-Pospelov Model}
In this section we propose a generalization of 
the fermionic P-U model
to a quantum field theory. This generalization 
may be 
performed in different ways.
However, we will use the following approach: we consider the 
standard field theory as an ensemble of decoupled 
harmonic oscillators in the momentum 
space, with and additional term containing higher 
time derivatives, as we did in the previous section.
\subsection{Covariant free field theory as ensembles of harmonic oscillators}
To fix ideas, let us review how we can 
go from the standard 
bosonic and fermionic harmonic oscillator,
to a scalar and Dirac free field theories, 
respectively. 

First, we consider an ensemble of bosonic harmonic 
oscillators whose dynamical variables
are labeled by a vector $\vec p$ with frequency 
$\omega_{\vec p}$, so that the Lagrangian 
of this ensemble is a sum of the Lagrangian of the 
individual oscillators over all possible 
values of ${\vec p}$. In the continuum limit the sum becomes 
an integral, and we can write
the Lagrangian as
\begin{eqnarray}
L=  -\frac{1}{2} \int d^3p \, \phi^{\dag}_{\vec p}
(\partial_t^2+\omega^2_{\vec p} ) \phi_{\vec p}
\end{eqnarray}
where $\phi_{\vec p}$ is the dynamical variable of 
the oscillator.
In order to make contact with a relativistic theory we 
assume that $\omega_{\vec p}=\sqrt{\vec p^2+m^2}$.
After a Fourier transform we end up with the free 
scalar field theory  
\begin{eqnarray}
L= -\frac{1}{2} \int d^3x  \phi_{ \vec x}
( \Box+m^2)   \phi_{ \vec x},
\end{eqnarray}
where $ \phi_{ \vec x}(t)$ is the Fourier 
transform of $ \phi_{ \vec p}(t)$. 

The fermionic case is less direct. In order to 
recover the free Dirac
field,  we will need four fermionic oscillators at each 
point in the $\vec p$ vector space.
We will label this fermion variables as 
$\psi_{i \vec p}^s $ where $s$ will be the spinor 
index and $i$ will specify
 particle or antiparticle modes. Then, the 
Lagrangian at each point $\vec p$ is 
\begin{eqnarray} \label{HOT}
L_{ \vec p}= \sum _ {s,i}  \psi_{ i\vec p }^{\dag s} 
(i\partial_t-\omega^s_i(\vec p))\psi_{ i\vec p }^{s}.
\end{eqnarray}
Again, the ensemble of this oscillators will be given 
through the integral over all possible values of 
$\vec p$. To get the relativistic 
theory, we assume that the frequencies 
have the form 
\begin{eqnarray} \label{HOT1}
\omega^s_i(\vec p)= \epsilon_i \omega_{\vec p},  
\end{eqnarray}
with $\epsilon_i=\pm 1$ for particles and antiparticles,
 respectively, and we write the new variables
\begin{eqnarray} \label{Var}
 \psi_{\vec p}(t)= \frac{1}{\sqrt {2 \omega_{\vec p}}} 
  \sum_{s,i} 
 w_i^s(\vec p) \psi_{i \vec p }^s(t),
\end{eqnarray}
where $w_1^s (\vec p )= u^s(\vec p )$ is the amplitude of the free
 wave function that is solution of the Dirac equation for particles of spin $s$
and momentum $\vec p $ and  $w_2^s (\vec p )= 
v^s(-\vec p )$ the free wave function for 
antiparticles of spin $s$
and momentum $-\vec p $. In terms of these
basis and making use of the orthogonality relation 
\begin{eqnarray}\label{orthogonality}
w ^{\dag r}_{i} ( \vec p) w ^s_{j}(\vec p)  
 = 2\omega_{\vec p} \delta^{rs}  \delta^{ij},
\end{eqnarray}
and the completeness relations 
\begin{eqnarray} \label{completeness}
\sum_s{w^s_i}(\vec p) w^{s\dag}_i(\vec p)= 
\omega_{\vec p} + \epsilon_i  h_D(\vec p) ,
\end{eqnarray}
where $ h_D (\vec p)=\vec \alpha \cdot
 \vec p +m\beta$ and $\vec \alpha=\gamma^0\vec{\gamma}$ and $ \beta=\gamma^0$ are
the $4\times 4$ Hermitian Dirac matrices, the Lagrangian turns out to be
\begin{eqnarray}
L_{ \vec p}= \psi^{\dag}_{ \vec p}(i\partial_t
-h_D(\vec p) \psi_{ \vec p}.
\end{eqnarray}
Hence, the Lagrangian of the ensemble, which is 
the integral of the above expression, 
can be written in terms of the Fourier 
transform variables 
$\psi_{ \vec x}$ and $\bar \psi_{ \vec x}=
\psi^{\dag}_{ \vec x}\beta $ as
\begin{eqnarray}
L= \int \frac{d^3p}{(2\pi)^3} L_{ \vec p}= 
\int d^3x \bar 
\psi_{ \vec x}(i\partial \cdot \gamma-m) \psi_{ \vec x}.
\end{eqnarray}

Summarizing, this procedure shows us that 
given an ensemble of oscillators, 
with frequencies depending on free parameters, 
we can find a space
that we call $x$-space where the ensemble is 
described as a relativistic 
standard free field theory. 
\subsection{An ensemble of P-U harmonic oscillators: Fermionic Myers-Pospelov model}
Now, we come to the main part of the section. 
As it was mentioned before, we will consider 
an ensemble of P-U harmonic oscillators and 
see what theory produces. 

We set out from Eq. (\ref{HOT}) and (\ref{HOT1}), but we add a 
higher time derivative term
\begin{eqnarray}\label{var2}
L_{ \vec p}= \sum _ {s,i}  \psi_{ i\vec p }^{\dag s} 
(g \partial_t^2+i\partial_t-  \epsilon_i 
\omega_{\vec p}) \psi_{ i\vec p }^{s}.
\end{eqnarray}
Then we consider the variable $\psi_{ \vec p }$ 
as defined in Eq. (\ref{Var}) 
and performing a Fourier transform the Lagrangian 
has the form 

\begin{eqnarray}
L= \int \frac{d^3p}{(2\pi)^3} L_{ \vec p}= 
\int d^3x \bar 
\psi_{ \vec x}(i\partial \cdot \gamma+g \partial^2_t- m)
 \psi_{ \vec x}.
\end{eqnarray}
This model, arising from an ensemble of P-U harmonic 
oscillators, can be identified 
with fermionic sector of the well known M-P model, 
which is an effective Lorentz invariance violating field 
theory, containing quantum gravity effects.
To clarify this claim, we introduce a vector $n=(1,0,0,0)$, 
and the former Lagrangian can be written as
\begin{eqnarray}
L= \int d^3x \bar \psi_{ \vec x}(i\slashed\partial+g \slashed 
n(n\cdot\partial)^2- m) \psi_{ \vec x}.
\end{eqnarray}
This is the Myers and Pospelov fermionic sector with $g_2=0$
and an isotropic background with 5-dimension operators. This model 
has been widely studied in the context of quantum gravity 
effects \cite{MP,Lopez-Reyes}.
\section{Unitarity}
Indefinite metric theories have potential physical problems 
which may spoil out any predictive ability and even worse, 
they would lack of a sensible physical interpretation \cite{nonhermitian}. 
The source of these problems is the failure of the unitarity 
of the time evolution operator. 
Many years ago, Cutkosky proposed a prescription to keep the 
unitarity at perturbative level for nonhermitian but Lorentz 
symmetric theories \cite{INDFMETRIC}. This procedure was successfully performed 
for the so called Lee-Wick models.

In this section, we attempt to fit similar prescription in order to prove the unitarity for the M-P fermionic sector at one loop perturbative level, and for energies less than the ultraviolet scale of the theory. This modification must take into account the Lorentz invariance violation intrinsic in the M-P model. As it was mentioned in the introduction, we are only interested in this low energy regime because it is the relevant one for a theory which must be seen as  an effective theory parametrizing new features of  space-time structure which may come from a more fundamental theory.
\begin{figure}
\centering
\includegraphics[width=0.47\textwidth]{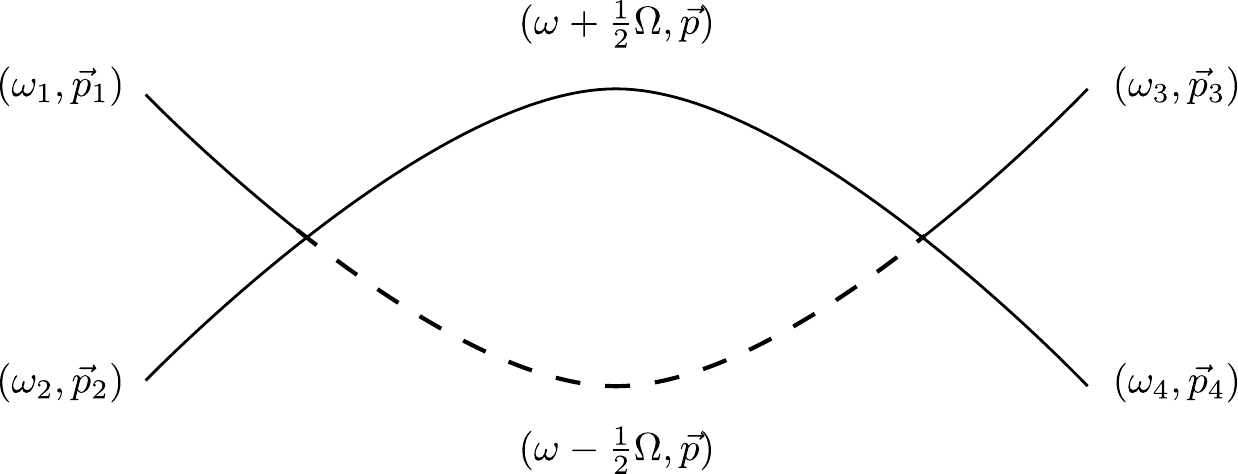}\vspace{15pt}
\includegraphics[width=0.47\textwidth]{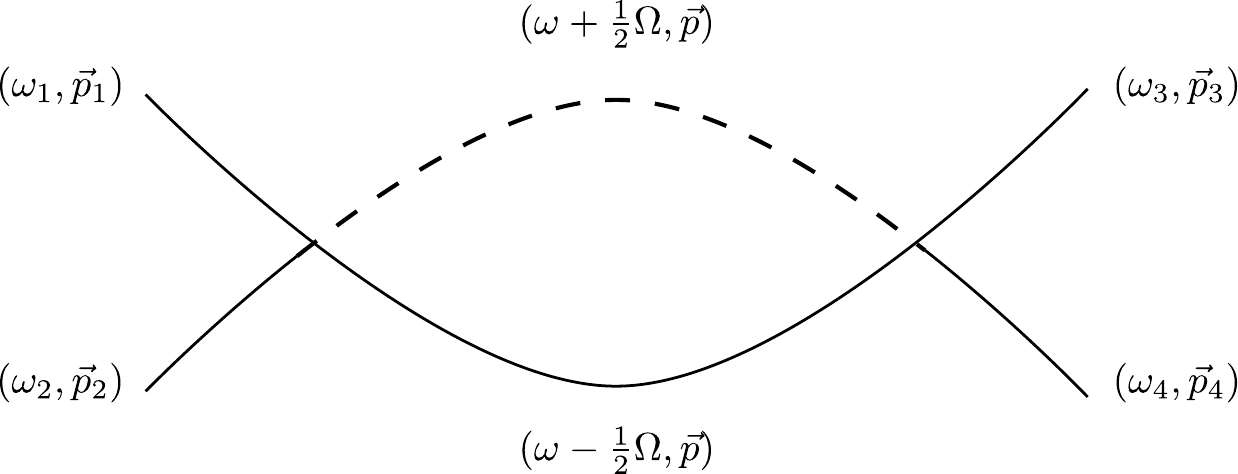}\vspace{15pt}
\includegraphics[width=0.47\textwidth]{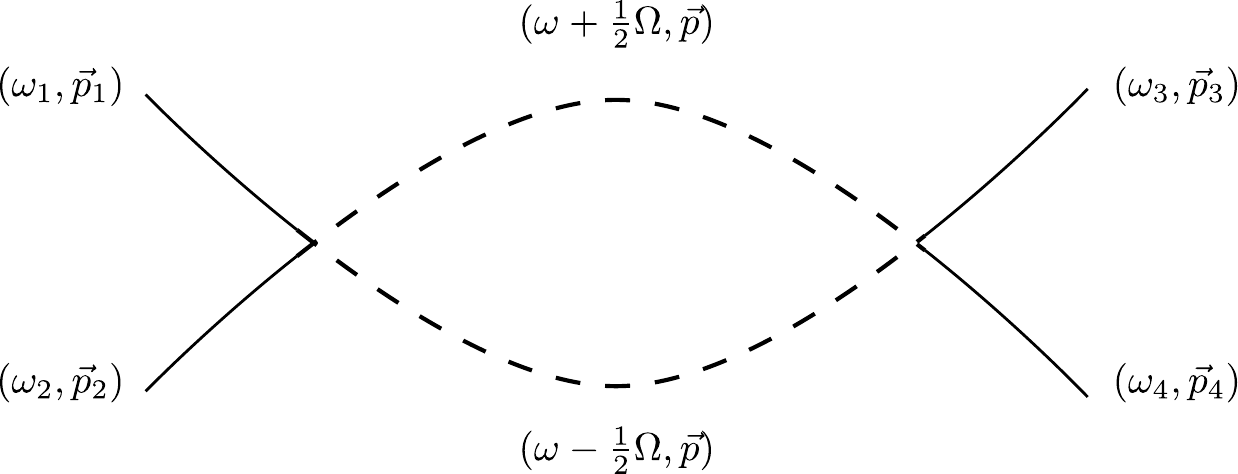}
\caption{We show the diagrams that contain ghosts and therefore those that 
can spoil unitarity at one loop order. In the first figure we show the diagram 
corresponding to the amplitude $M_{+-}$, in the second, the amplitude $M_{-+}$,
and in the last figure the amplitude corresponding to $M_{--}$. }
\end{figure}\label{fig1a}	

The first thing to do is to write down the relevant propagators.
In the M-P theory the propagators, in terms of fundamental
fields are complicated, and it is more convenient to 
express them in terms of the P-U modes. These are the P-U modes
\begin{eqnarray}\label{psi}
\psi^{r,i}(t,\vec p) \equiv 
\frac{1}{\sqrt{2\omega_{\vec{p}}}}\,   w^{r\dag}_i  (\vec p)\psi(t,\vec p),
\end{eqnarray}
in terms of the Fourier transform fields \\ $\psi(t,\vec p)
=\int d^3x \psi(t,\vec x)e^{i\vec p\cdot \vec x}$.
Now these new variables correspond to four P-U oscillators 
with frequency
$\omega(\vec p)=\epsilon_i \omega_{\vec p}$. As we have seen in section 
II this oscillators are equivalent to two 
fermionic oscillators with frequencies 
\begin{eqnarray}
\omega_{\pm i}^r(\vec p) \equiv \frac{1\mp\sqrt{1-4g \epsilon_i 
\omega_{\vec p}}}{2g},  
\end{eqnarray}
for each P-U oscillators labeled with $i$ and $r$. The $\pm$ 
signs refers to the signs of the norm 
for those modes. Now, it is straightforward to calculate the 
propagators,
\begin{eqnarray}
G_{Fab }^{ rs,ij}(t,\vec x) \equiv 
\langle 0   \vert  {\mathcal T}  (\psi^{r}_{ia } (t,\vec x)
\psi^{s\dag}_{jb } (0)   \vert  0\rangle,  
\end{eqnarray}
corresponding to these modes by looking at the expression 
at the end of the section II. Namely the Fourier transform 
of this propagators are,
\begin{eqnarray}\label{green2}
G_{Fab }^{ rs,ij}(\omega,\vec p) = a\,  \frac{i}{2
\omega_{\vec p}} \times \frac{ w^r_i(\vec p)w^{s\dag}_j(\vec p) } 
{\omega-\omega^r_{a i}(\vec p)+i\delta^a_i }\delta_{ab}, 
\end{eqnarray} 
where $\delta^a_i$ will correspond to some suitable
prescription to fit later. In order to recover the 
standard Lorentz symmetric limit we must impose the 
prescription $\delta^+_1= -\delta^+_2=\epsilon >0$. To see this, 
the  standard   Feynman  propagator  can  be  related   with  the  P-U
propagators by 
\begin{eqnarray}
\Delta_F(\omega, \vec p)= \sum_{s,i}G_{F++ }^{ ss,ii}(\omega,
\vec p) \gamma^0.
\end{eqnarray}
Then, in the limit $g \to 0$ we have $\omega_{+i}^s (\vec p) \to 
\epsilon_i\omega_{\vec p}$, 
and making use of the completeness relations (\ref{completeness}) 
we obtain the standard relativistic expression
\begin{eqnarray}
\Delta_F(\vec p)&=& \frac{i}{2\omega_{\vec p}}\left[\frac
{\omega_{\vec p}+ h_D(\vec p)}{\omega-\omega_{+1}
(\vec p)+i\epsilon} 
+\frac{\omega_{\vec p}- h_D(\vec p)}{\omega-
\omega_{+2}(\vec p)-i\epsilon}  \right]\gamma^0\nonumber \\
&\to&\frac {i(\sla{p}+m)}{p^2-m^2+i\epsilon}.
\end{eqnarray}
Now, we will prove that with a suitable prescription for $\delta^-_i$
we can keep the unitarity of the theory at one loop perturbation, 
at least. Let us write the evolution operator in terms of the 
scattering matrix,
\begin{eqnarray}
\mathcal U= 1+iS.
\end{eqnarray}
Then, the unitarity implies,
\begin{eqnarray}
\mathcal U \mathcal U^\dag = 1+i(S-S^\dag )+SS^\dag =1.
\end{eqnarray}
Namely, we must prove that $-i(S-S^\dag)= SS^\dag$.

Perturbatively, we must show that
\begin{eqnarray}\label{unitarity}
2{\rm {Im}}\, \mathcal M({\rm{in\vert out}})= \sum_{\rm{Phys}} 
\mathcal M(\rm{in\vert Phys})\mathcal M(Phys\vert out),
\end{eqnarray}
where $\mathcal M$ are the perturbative elements of the scattering 
matrix and the sum is over all possible physical states  allowed by 
the conservation of energy and momentum.

We are interested in showing this equality for in and out states 
for particles with energies less than the ultraviolet scale $\frac{1}{4g}$.
At tree level, due to the energy and momentum conservation, it is 
impossible to create ghosts states and then, the contribution is 
zero at both sides of the previous equation, at first order in 
perturbation correction.  The first chance to break down the 
equation (\ref{unitarity}) is at one loop order; however, for the 
same reason above mentioned, the right hand side of equation  
(\ref{unitarity}) is zero. Then, we must check that the imaginary 
part of $\mathcal M$ is zero at the first radiative correction.
	
It is sufficient to analyze the integrals corresponding to the diagrams 
in the Fig.\ 1, which have problematic ghost internal lines. 
To simplify, we will work in the center of mass frame,
with total energy $\Omega$ and total momentum $\vec{P}=0$. 
The integrals contributing to this graphs are, 
\begin{eqnarray}
\mathcal M_{ab}(\Omega)= \frac{\alpha^2 }{2}  
\sum _{\overset {r,s,r',s'} {i,j,i',j'}}  
 \int d\omega  d^3 p \nonumber \\   {\rm{tr}}\left
 (G_{Faa}^{ rs,ij}(\omega-\Omega/2,\vec p) 
G_{Fbb }^{ r's',i'j'}(\omega+\Omega/2,\vec p)\right) 
\end{eqnarray} 
where $\alpha$ is the coupling constant which 
introduces an interaction and $a,b=\pm$. 

By using the orthogonality
property in the Eq. (\ref{orthogonality}) the sum 
of the integrands  of these diagrams are  proportional to the functions,
\begin{eqnarray}
\Gamma_{ab}(\omega)
=\sum_{i}&&\frac{1}{\omega-\Omega/2-\omega_{ai}(\vec{p})+i\delta_i^a
} \times \nonumber
\\
&&\frac{1}{\omega+\Omega/2-\omega_{bi}(\vec{p})+i\delta_i^b  }
\end{eqnarray} 
The pole structure 
of these integrands is shown in Fig.\ 2 and Fig.\ 3, and their residues at the poles
are
\begin{eqnarray}
\mathop{\rm Res}_{\omega_{ai}-\Omega/2}\Gamma_{ab} &=& 
\frac{1}{\omega_{ai} -\omega_{bi}-\Omega},\\
\mathop{\rm Res}_{\omega_{bi}+\Omega/2}\Gamma_{ab} &=& 
\frac{1}{\omega_{bi} -\omega_{ai}+\Omega}.
\end{eqnarray}
 Any prescription to avoid 
the poles is equivalent 
to a path integration in the $\omega$-plane $\mathcal C$. 

For   $\omega_{\vec  p}<\frac{1}{4g}$,   we  choose   the  prescription
$\delta^-_1=-\delta^-_2=\delta  >0$. When  we integrate  $\omega$ over
the real axis, this prescription leave the poles $1^-$ below and $2^-$
above.  This  is  equivalent  to  the  contour  ${\cal  C}$  shown  in
Fig. 2. The conjugate of this integration is given by the contour
${\cal  C}^*$.   Hence,  the  imaginary  part  of   the  integral  is
proportional  to   the  closed  contour   integration  over  $\mathcal
C-\mathcal C^*$, which is proportional to the sum of the residues.

Thus, for ${\rm {Im}} \mathcal M_{--}$ we have,
\begin{eqnarray}
\oint_{{\cal C}- {\cal C}^*} \Gamma_{--}\, d\omega & \propto & \sum_{\mbox{\tiny{Poles}}}
\mathop{\epsilon_i\rm Res}_{\omega_{-i}\pm\frac{\Omega}{2}}
\Gamma_{--} (\omega),
\nonumber\\
&=& \sum_i\epsilon_i\left(\frac{1}{-\Omega} + \frac{1}{\Omega}\right) =0.\nonumber
\end{eqnarray}
And for 
${\rm {Im}} (\mathcal M_{+-}+ \mathcal M_{-+})$, we have
\begin{eqnarray}
\oint_{{\cal C}- {\cal C}^*}&& (\Gamma_{+-} +\Gamma_{+-}) d\omega  \propto \sum_{\mbox{\tiny{Poles}}}
\mathop{\epsilon_i\rm Res}_{\omega_{\pm i}\pm\Omega/2}
(\Gamma_{+-} + \Gamma_{-+}),\nonumber
\\
&&=  \sum_{i}\epsilon_i \left( \frac{1}{\omega_{+i}-\omega_{-i}-\Omega}
+\frac{1}{\omega_{-i}-\omega_{+i}+\Omega}\right)=0,\nonumber
\end{eqnarray}
where $\epsilon_i$ stands for signs $\pm 1$ of the closed contours orientation that surround the poles.

For internal momentum above the critical 
scale,  $\omega_{\vec p}>\frac{1}{4g}$, some poles become complex, as it can be seen 
in Fig.  3. Now  the integral over  the real  axis of $\omega$  can be
closed around the lower half-plane, picking up the residues at the
poles in this  half-plane.  The complex conjugate  integration will be
closed above  the upper-half plane,  picking up now the  poles in this
half-plane.  Thus, the  imaginary part  of the  integral  with complex
poles is given  by a contour enclosing these  poles. This prescription
is equivalent to 
take the poles corresponding to $1^+$ moving 
up and the ones corresponding to $1^-$
moving down. Then, it can be seen that the 
prescription given above corresponds to take the integration path
$\mathcal C$ and $\mathcal C^{*}$ shown in Fig. 3, 
which are smooth deformations 
of those paths in Fig. 2.

Now, the closed contour integrals take the residues in the same way as
before, and their sum vanishes again.

Thus, we have found a particular prescription 
to maintain the
unitarity at one loop order. 
\begin{figure}
\centering
\includegraphics[width=0.47\textwidth]{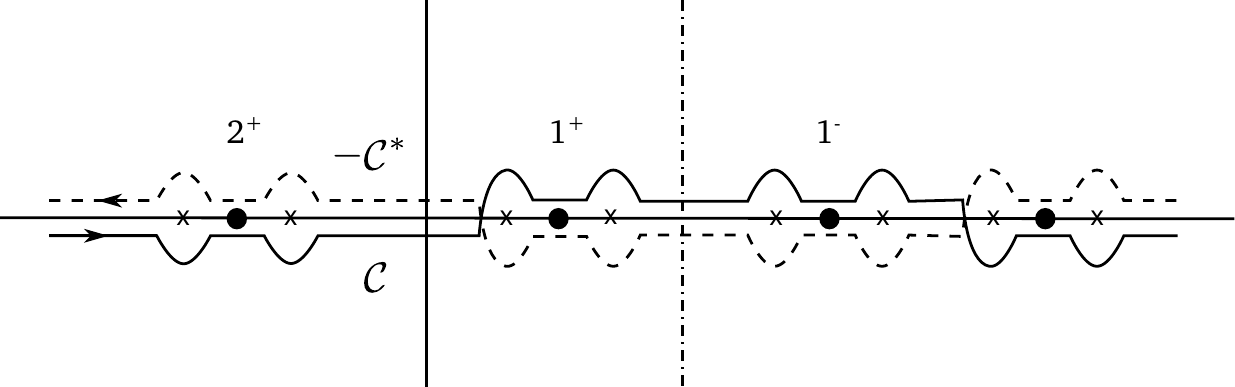}
\caption{We show the pole structure for $\omega_{\vec{p}}<1/4g$, 
the notation $i^{\pm}$ stands for particles
and antiparticles with positive or negative norms ($\pm$). 
The crosses correspond to the shifted poles
in the integrals of the amplitudes at one loop order. 
The contour $C$ corresponds to the prescription described
in the text,
$C^*$ to its conjugate. The vertical line corresponds to 
the critical region above which 
the poles become complex.}
\end{figure}\label{fig2}
\begin{figure}
\centering
\includegraphics[width=0.47\textwidth]{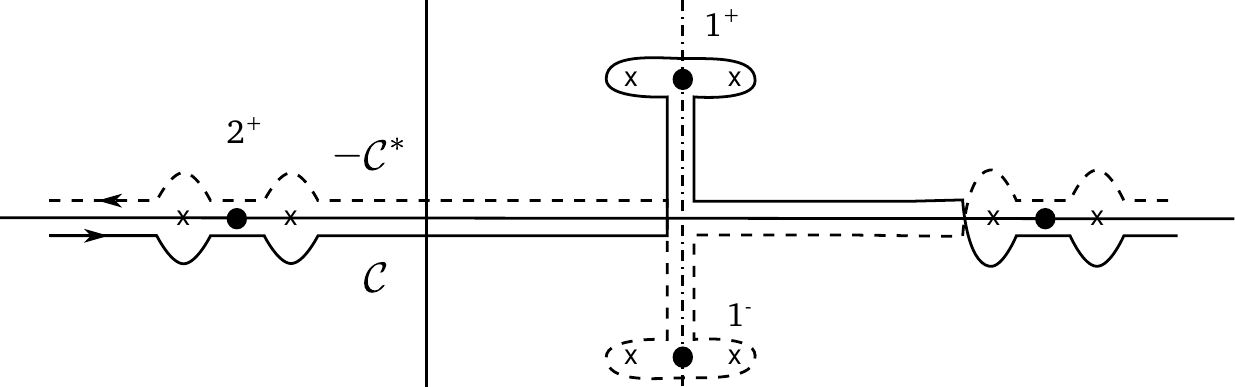}
\caption{The pole structure above the critical 
region $\omega_{\vec{p}}>1/4g$. The poles corresponding to 
$1^+$ and $1^-$
collapse to each other and they become complex. According to our 
prescription the pole $1^+$ go up and $1^-$ go down
in the critical vertical line. The integration contours 
$C$ and $C^*$ are smooth deformation of those of the Fig. 2.}
\end{figure}\label{fig3}	

We have used a somehow
arbitrary interaction, however, the procedure 
followed above can work with
other general interactions.

Finally, let us note that there is some 
ambiguity in choosing the
prescription which keep the unitarity.  
This freedom could be 
used to maintain the unitarity at higher 
radiative corrections.
Nevertheless, this is far from being clear 
in this analysis.
\section{Final Comments}
Summarizing, in this work we have found a 
connection between 
an emsemble of fermionic P-U oscillators, and the well 
known M-P model in the fermionic sector. This fact has allowed us to prove  the unitarity of the M-P fermonic sector at one loop order in perturbation theory at low energies where the theory is applicable, even though, it contains higher time derivative terms. 
This 
statement is the most important result in this 
study, because 
it permits to have sensible predictions and 
physical interpretations.  

Our study shows that for Lorentz invariant violating theories with higher time derivative 
terms -- with fermions -- it is possible to maintain the 
unitarity at first order in radiative corrections. 

The prove of the unitarity at higher 
radiative corrections, 
is left as a future work.  Also, it would be interesting to 
do phenomenological predictions from this theory.

\begin{acknowledgements}
We thank T. Fernandez and Prof. D. Restrepo for helpful support during
the preparation of 
 the manuscript.
J.L-S. acknowledges support from DICYT Grant 
No. 041131LS (USACH) 
and FONDECYT-Chile Grant No. 1100777 and wants to thank the 
hospitality at the Universidad de Antoquia.  
C. M. R. acknowledges
partial support from 
Direcci\'on de Investigaci\'on de
la Universidad del B\'{\i}o-B\'{\i}o (DIUBB) 
Grant No. 123809 3/R.
\end{acknowledgements}
\appendix \label{appendix}
\section{Schr\"oedinger Quantization of the P-U Fermionic Oscillator}
In this appendix we show an alternative procedure to 
quantize the P-U fermionic oscillator. 
This formalism gives us an explicit form of the vacuum 
state, and shows how the negative norm states appear.

The following Hamiltonian 
\begin{eqnarray}
H= -g\dot{\bar\psi}\dot\psi + \omega\bar\psi\psi,
\label{hamil}
\end{eqnarray}
and the non vanishing anticommutators
\begin{eqnarray}
\{\dot {\bar \psi},  \psi\}= -\frac{i}{g},
\quad
\{\dot  \psi, \bar  \psi\}= \frac{i}{g},\label{atc1}
\quad
\{\dot \psi,  \dot {\bar \psi}\}= -\frac{1}{g^2}. \label{atc2}
\end{eqnarray}
reproduce the fermionic P-U equations of motion presented in section II.

Using the Schr\"oedinger representation in terms of Grassman 
variables and their derivatives, we have,
\begin{eqnarray}
\dot {\bar \psi}&=&-\frac{i }{g }\frac{\partial }{\partial 
\psi}+\frac{i}{2g}\bar \psi,\label{psidot}\nonumber \\
\dot { \psi}&=&\frac{i }{g }\frac{\partial }{\partial 
\bar\psi}-\frac{i}{2g} \psi.\label{barpsidot}
\end{eqnarray}
And the Hamiltonian in this representation is,
\begin{eqnarray}
:H:&=&-\frac{1 }{g}\frac{\partial }{\partial \psi}\frac{\partial }
{\partial \bar \psi}+\omega\left(1-\frac{1}{4g\omega}\right)\bar \psi\psi
\nonumber\\
&+&\frac{1} 
{2g}\left( \bar \psi \frac{\partial }{\partial \bar \psi} -\psi
 \frac{\partial }{\partial \psi}\right)+\frac{1}{2g}.
\label{schrhamil}
\end{eqnarray}
The first and second lines of this equation commute to each other, so it is very 
easy to find the eigenfunctions,
\begin{eqnarray}
\begin{array}{lcl}
\Phi_0=\psi, & & E_0 = 0, \label{0} 
\\
\Phi_1   =   \frac{\sqrt  2}{(1-4g\omega)^{\frac{1}{4}}}\,   e^{\frac{
    \sqrt{1-4g\omega}}{2}\, \bar\psi\psi},& & E_1=\omega_+,\label{1} 
\\
\Phi_2= \frac{\sqrt 2}{(1-4g\omega)^{\frac{1}{4}}}e^{-\frac{\sqrt{1-4g\omega}}{2}
\,\bar\psi\psi}, & & E_2 = \omega_-,\label{2}
\\
\Phi_3=\bar\psi, & & E_3 = \omega_{+} + \omega_-, \label{3}
\end{array}
\end{eqnarray}
where $\omega_{\pm}$ are defined in section II. For positive $g$, 
the state $\Phi_0$ corresponds to the lowest energy state, and 
then it represents the vacuum of the theory. 
%
%
Now, it is clear, that this is a four state system with bounded energies.

Regarding the normalization, we can define the Berezin measure as 
$\int d\bar\psi\,d\psi\, \bar \psi\psi= +1$, the scalar product  of wave functions
$F(\psi,\bar\psi)$ and $G (\psi,\bar\psi)$ is 
$$
\langle G\vert F\rangle\equiv \int d\bar\psi\,d\psi\, \bar G F.
$$
Then we can see that
\begin{eqnarray}
\langle\Phi_0\vert\Phi_0\rangle =
\langle\Phi_1\vert\Phi_1\rangle=1. 
\end{eqnarray}
However,
\begin{eqnarray}
\langle\Phi_2\vert\Phi_2\rangle=\langle\Phi_3\vert\Phi_3\rangle=-1 
\end{eqnarray}
are states with negative norm. This agrees with the results in section II.

\end{document}